**Identification of post-COVID-19 symptoms using brain structural MRI features: a machine learning approach**

Abdi Reza [a,b]

[a] Department of Neurosurgery, Faculty of Medicine, Universitas Indonesia, DKI Jakarta, Jakarta 10430, Indonesia.

[b] Department of Neurosurgery, RSUP Nasional dr. Cipto Mangunkusumo, Jl. Pangeran Diponegoro No. 71, Jakarta Pusat, DKI Jakarta, Jakarta 10430, Indonesia.

**Abstract**

*Background*. Identifying long-COVID symptoms is a challenging task, primarily due to the reliance on patient reports and the lack of disease-specific biomarkers. Brain structural magnetic resonance imaging (MRI), which has been increasingly employed for the comprehensive evaluation of patients with COVID-19, offers the neuroimaging biomarkers and the opportunity to identify the post-COVID-19 symptoms. The objective of this study is to identify individual long-COVID symptoms, post-COVID-19 conditions (PCC) participants, and participants' sex, and to identify the associated brain regions by developing an explainable machine learning algorithm using brain MRI features.

*Methods.* This study implements secondary analysis using an anonymized, publicly accessible dataset that categorizes participants into three groups: the PCC group, the Unimpaired Post-COVID-19 group (UPC), and the Healthy Non-COVID group (HNC), each with corresponding symptoms, demographics, and brain structural MRI features. The aim is to develop and cross-validate a support vector classifier (SVC) algorithm to identify the occurrence of various target labels, including specific symptoms (e.g., 'word finding difficulties', 'multitasking'), participants' group (PCC or non-PCC), and sex-male from the dataset.

*Results*. Among PCC and UPC participants, the SVC classifier successfully identified the occurrence of PCC symptoms of 'word finding difficulties' with cross-validated balanced accuracy of 68.375 $\pm$ 8.614 % ($P$ = 0.001; permutation test), 'multitasking' with cross-validated balanced accuracy of 67.279 $\pm$ 12.259 % ($P$ = 0.001; permutation test). The PCC group can still be distinguished from PCC and UPC participants with cross-validated balanced accuracy of 62.202 + 9.785 % ($P$ = 0.013; permutation test). However, the cross-validated balanced accuracy dropped when all participants (PCC, UPC, and HNC) were combined (balanced accuracy of 55.425 ± 5.766%; P = 0.125; permutation test). Nonetheless, the male-sex participants are still successfully identified with a cross-validated balanced accuracy of 81.786 ± 6.209 % ($P$ = 0.001, permutation test) from all PCC, UPC, and HNC participants.

*Conclusion*s. The brain structural MRI contains invaluable information associated with changes due to post-COVID-19. The ML task of identifying specific PCC symptoms among COVID-19 survivors is more favorable than distinguishing the overall PCC participants. Considering the current reliance on patient reports for diagnosing long-COVID symptoms, the demonstrated approach offers an alternative modality for identifying the occurrence of long-COVID symptoms based on neuroimaging biomarkers.

**Identification of long-COVID symptoms using brain structural MRI features: a machine learning approach**

1. Introduction

COVID-19 has caused a global catastrophe since its massive spread in 2019 (Ciotti et al., 2020). Moreover, various prolonged symptoms such as olfactory dysfunction, cognitive deficit, and fatigue, which are later described as "post-COVID-19 conditions" (PCC) (Soriano et al., 2022) or so-called "long-COVID" (Unger, 2025), could still impact many people with COVID-19, even after the infection period (Augustin et al., 2021). Although the population cohort shows that the severity of COVID-19 infection is generally mild (Sun and Yeh, 2020; Wang et al., 2025), the PCC may still occur among patients with mild symptoms (Augustin et al., 2021; Boscolo-Rizzo et al., 2021; Fernández-Castañeda et al., 2022; Lai et al., 2023). Subsequently, it downgrades the quality of life (QoL) (Tabacof et al., 2022).

Diagnosing patients with long-COVID symptoms is a challenging task, potentially due to the reliance on the patient's report (Kisiel et al., 2023; O'Hare et al., 2022), unreported past asymptomatic/mildly symptomatic patients (Raveendran, 2021), confusing false negative findings of the Reverse Transcription Polymerase Chain Reaction (RT-PCR) test (Raveendran, 2021), possible COVID-19 effects on cognitive function and communications (Cummings, 2024), and a lack of disease-specific biomarkers (O'Hare et al., 2022). Therefore, an alternative screening method that is independent of patient report and based on an objective biomarker is needed. Currently, researchers have attempted to identify patients who develop PCC, starting with identifying the risk factors (Reme et al., 2023; Zhao et al., 2023) to propose a machine learning (ML) model (Lo et al., 2025; Pfaff et al., 2022; Reme et al., 2023). Since an improved recognition could be paired with corresponding treatments for specific long-COVID symptoms (Pettemeridou et al., 2025; Zeraatkar et al., 2024), a decrease in the QoL could be prevented.

Brain magnetic resonance imaging (MRI) has been increasingly utilized in COVID-19 evaluation, particularly for those who developed neurological symptoms (Radmanesh et al., 2020; Tan et al., 2021), under intensive care treatment (Kandemirli et al., 2020), or are undergoing follow-up (Du et al., 2023; Hellgren et al., 2021; Pihlajamaa et al., 2025). Beyond neuroanatomical diagnostic purposes, brain MRI also provides an opportunity to examine the brain structures that carry specific functions; it is well-known that by utilizing brain MRI features, ML algorithms could predict physiological labels such as gender (Wiersch et al., 2023) and age (Baecker et al., 2021), as well as pathological labels such as autism spectrum (Nogay and Adeli, 2024), attention-deficit/hyperactivity disorder (Tian et al., 2024), and cognitive impairment (Zubrikhina et al., 2023). Unsurprisingly, the utilization of brain MRI has revealed abnormalities associated with COVID-19 infections (Wagner et al., 2023). However, despite the emerging literature showing associations between MRI brain findings and anosmia, cognitive disturbances, and fatigue (Heine et al., 2023; Hellgren et al., 2021; Lu et al., 2020) among post-COVID-19 patients, brain MRIs have rarely been utilized to point out the occurrence of long-COVID symptoms. It remains unclear which long-COVID symptoms can be identified from brain MRI features and which brain MRI features are responsible for corresponding symptoms. Considering the obstacles to long-COVID diagnosis (Raveendran, 2021) and past experience of COVID-19 reemergence (Lemey et al., 2021; Menhat et al., 2024), the brain MRI's potential contribution to identifying long-COVID symptoms needs to be explored, as it is widely available, provides an objective neuroimaging biomarker, and is independent of patient reports.

This study hypothesizes that brain structural MRI can identify long-COVID symptoms, including the presence of fatigue, "finding-word-difficulty", attention, memory, multitasking problems, and acute and chronic olfactory symptoms, and separate the PCC population from non-PCC populations as well. Additionally, sex-male identification from brain MRI features is implemented to confirm the replicability of previous literature. To test this hypothesis, a machine learning algorithm is trained using a publicly available dataset from a prior study (Hosp et al., 2024) that reported post-COVID-19 conditions and the brain MRI features of patients. The dataset is referred to as the "Freiburg-MRI" dataset throughout this manuscript. The publicly anonymized dataset provides an opportunity to develop the model without the need to collect new data prospectively, ensuring transparency and replicability. The model performance and brain MRI features are also examined to determine their impact on model prediction and are discussed in light of previous research.

## 2. Methods

This is a proof-of-concept study that ML algorithms could identify the presence or absence of long-COVID symptoms using brain MRI features. To achieve this aim, a secondary analysis is conducted on a previously published prospective cohort study that investigated changes in microstructural brain MRI features following COVID-19 infections. The original research provides detailed information on participant inclusion and exclusion, data collection, measurement, feature extraction methods, and the ethical review process (Hosp et al., 2024). Ethical review for the source study was approved by the Ethics Committee of the University of Freiburg (EK 211/20). It was registered in the 'Deutsches Register Klinischer Studien (DRKS)' (DRKS00021439), adhered to the Declaration of Helsinki, and all participants provided written informed consent for study participation. The publicly available dataset has been anonymized, and it can be accessed from the Dryad data center (Hosp et al., n.d.), available from https://doi.org/10.5061/dryad.kkwh70s9g, no further ethical review is necessary for the secondary analysis of the anonymized dataset, which is distributed under a public domain license (https://creativecommons.org/publicdomain/zero/1.0/).

### 2.1. Dataset description and preparation

In short, the participants consist of three groups: first, the Post-COVID-19 Condition (PCC) group. Second, the Unimpaired Post-COVID group (UPC). Third, the Healthy Non-COVID (HNC) group. The PCC participants are recruited based on findings of a SARS-CoV-2 infection confirmed by reverse transcription polymerase chain reaction (RT-PCR); fulfillment of diagnostic criteria for Post-COVID-19 Condition according to WHO criteria (Soriano et al., 2022), and are eligible for cranial MRI. The UPC participants are recruited from patients in the chronic phase following PCR-confirmed COVID-19 infection, who do not have persistent subjective complaints. The HNC participants are those with no history of COVID-19 infection, as documented in medical records and self-reports, and have no significant age difference compared to participants in the PCC and UPC groups. Exclusion criteria were any pre-existing neurodegenerative disorder, age below 18 years, and MRI artifacts.

The participants are evaluated by a neurologist to identify the occurrence of post-COVID-19 symptoms. Column 'attention' (attention deficit), 'memory' (memory deficit), 'multitasking' (multitasking disturbances), 'word_finding_difficulties', 'fatigue' (fatigue symptoms), 'olf_imp_ini' (acute olfactory symptoms), and 'olf_imo_late' (chronic olfactory symptoms) was given binary score '1' if corresponding symptoms are identified and '0' for non- finding. Non-binary evaluation (Hosp et al., 2024), is also provided in the anonymized dataset (Hosp et al., n.d.).

The binary label of PCC was evaluated to find missing values. The columns 'attention', 'memory', 'multitasking', 'word_finding_difficulties', 'fatigue', 'olf_imp_ini' (acute olfactory symptoms), and 'olf_imo_late' (chronic/late olfactory symptoms) contained empty values corresponding to 'HNC' in the column 'group' ('HNC' that represented healthy participants do not provide COVID-19 symptoms as PCC and UPC participants with a history of COVID-19 infection). Furthermore, the occurrence of symptoms in the column 'attention' is identical to that in the column 'memory' and 'PCC' in the 'group' column. Therefore, each ML model is trained separately for label 'memory'/'attention'/'PCC' (that consist of identical participants), 'multitasking', and 'word_finding_difficulties' as representation of cognitive domain, 'fatigue', 'olf_imp_ini', and 'olf_imo_late' from the dataset from pooled PCC and UPC, without HNC participants.

Since there were no empty values in the columns 'group' and 'sex', the column 'PCC' was created with '1' to represent the occurrence of 'PCC' from the column 'group' and the rest as '0'. The column 'gender_male' is created with '1' to describe the occurrence of 'Male' and the other as '0' in the column 'sex'. Each ML model was then trained separately to identify the occurrence of PCC and male gender from the entire dataset.

All participants underwent brain MRI scanning using a 3 Tesla scanner (MAGNETOM Prisma, Siemens Healthcare, Erlangen, Germany) with a 64-channel head and neck coil. Several cortical morphometry and diffusion microstructure imaging (DMI) calculations were derived as features. Detailed information regarding preprocessing, measurement, and patient clinical evaluation is accessible from the primary publication (Hosp et al., 2024) and is stored in the public repository (Hosp et al., n.d.). From the spreadsheet 'Combined' of the distributed dataset, the occurrence of not-a-number (Nan values) in the feature column of interest is evaluated, from the column

'total_gm_extra' to the last column that would be used as features; there were no empty cells within those columns. For current analysis, the column name from the original dataset, which contains abbreviations, is then expanded and adjusted to enhance readability, following the Desikan-Killiany atlas (Alexander et al., 2019; Desikan et al., 2006), Destrieux's nomenclature (Destrieux et al., 2010), the 'README.md' file accompanying the dataset, and FreeSurferWiki (https://surfer.nmr.mgh.harvard.edu/fswiki).

### 2.2. Model development and validation

A nested cross-validation (CV) is employed to tune the ML algorithm and evaluate the model performance. The dataset is divided by the outer loop constructed from a 5-fold stratified CV. Each loop trains the dataset using four folds and tests it with the remaining fold, by keeping both partitions separate to avoid information leakage. Within the outer loop, the numerical columns from the training data and test data are normalized using the standard scaling method, based on the values from the four-fold of the training data.

A support vector classifier with a linear kernel is chosen as the algorithm of choice. Support vector classifiers are capable of performing binary classification tasks with limited datasets (Khondoker et al., 2016), a linear kernel is chosen to avoid overfitting. Hyperparameter 'C' consists of [$2^{-6}, 2^{-5}, 2^{-4}, 2^{-3}, 2^{-2}, 2^{-1}, 2^{0}, 2^{1}$] was tuned using GridSearchCV() using the 5-fold-stratified-cv as inner-loop to determine the model with best performance, with balanced accuracy selected as the target evaluation metric. Other metrics, including Specificity, Accuracy, Matthew's coefficient, F1 score, Recall score, Precision score, Cohen's kappa score, and the ROC AUC are calculated as well. The features' influence was derived from the SVC coefficient, which shows the separability of the binary label by the closest point ("support vector") between the labels. The overall model performance and the feature coefficient were then derived from the average of the outer-loop model performances (Vieira et al., 2020).

## 2.3. Statistical procedures

To assess the significance of model performance, a permutation test is used to evaluate model performance by shuffling the target label 1,000 times using a unique randomization index in each iteration, followed by the nested CV process mentioned above. A thousand distributions of average model performance by chance level are derived from this process (Vieira et al., 2020). The occurrence of overall model performance from the null distribution of 1000-shuffling labels that is greater than or equal to the actual label was counted and added by one, then divided by 1001, as a correction factor (North et al., 2002). The alpha level of 0.05 is determined as the significance threshold. Due to multiple hypotheses from a single dataset, we also pay attention to alpha levels less than 0.05/8 to identify target labels with outstanding model performance. As a post hoc analysis, the same permutation test was also implemented to identify brain MRI features with an average coefficient lower than the alpha level of 0.05 (Vieira et al., 2020).

In general, numerical values are reported as the mean ± standard deviation (SD) whenever possible. The phi coefficient (Khamis, 2008), also known as the Matthews correlation coefficient (MCC) (Itaya et al., 2025), is used to evaluate the strength of association between binary variables. The upset plot is used to describe overlapping characteristics across participants (Lex et al., 2014).

## 2.4. Software

Python v.3.8.13 (Van Rossum and Drake, 2020). Several essential Python libraries are used to support data handling, analysis, model development, interpretation, and visualization. The libraries include Pandas v.1.4.2 (McKinney, 2010), Numpy v.1.21.4 (Harris et al., 2020), Scipy v.1.9.0 (Virtanen et al., 2020), Pingouin v.0.5.2 (Vallat, 2018), Statsmodels v.0.13.2 (Seabold and Perktold, 2010), Scikit-learn v.1.0.2 (Pedregosa et al., 2011), Shap v.0.39.0 (Lundberg and Lee, 2017), Matplotlib v.3.7.1 (Hunter, 2007), UpSetPlot v.0.9.0 (Lex et al., 2014), and Seaborn v.0.13.2,(Waskom, 2021).

## 3. Results

A total of 173 participants, comprising the PCC (Post-COVID-19 Condition; n = 89), UPC (Unimpaired Post-COVID group; n = 38), and HNC (Healthy non-COVID; n = 46) categories, are available for analysis. The participant inclusion flowchart is described in Figure 1. Readers are advised to refer to the source manuscript (Hosp et al., 2024) and the source dataset (Hosp et al., n.d.) for the inclusion process. A description of the participants' demographics across the three groups is presented in Table 1. Since the HNC participants did not report long-COVID symptoms, the prediction attempt using all participant observations is made only for the 'Male' and 'PCC' target binary labels, and the prediction of post-COVID conditions is made from pooled PCC and UPC participants (see Figure 2 for an overview of model development).

The Phi coefficient (also known as the Mathews correlation coefficient) is used to confirm the correlation between binary variables. From all categories, participants' 'Male' and 'PCC' show a phi coefficient of -0.0474. Among the pooled participants from UPC and PCC (Figure 3), the 'PCC', 'attention', and 'memory' show a phi coefficient of 1; they are considered to represent the same outcomes in this analysis (See also Supplementary Figure S1 for UpSetPlot for description of characteristics intersection across participants)

### 3.1. Model performance

The SVC model yields various performance results across different pooled datasets and target labels. Among the PCC and UPC participants the cross-validated balanced accuracy (%) results consists of acute olfactory symptoms (61.417 $\pm$ 6.059 (mean (%) $\pm$ standard deviation (SD))) ; $P$ = 0.025 ; $P < 0.05$), chronic olfactory symptoms (63.521 $\pm$ 11.408 ; $P$ = 0.011 ; $P < 0.05$), fatigue (62.639 $\pm$ 7.749 ; $P$ = 0.016 ; $P < 0.05$), attention/memory/PCC group (62.202 $\pm$ 9.785 ; $P$ = 0.013 ; $P < 0.05$), multitasking (67.279 $\pm$ 12.259 ; $P$ = 0.001 ; $P < 0.05/8$ with Bonferroni correction for multiple comparison), word finding difficulties (68.375 $\pm$ 8.614 ; $P$ = 0.001 ; $P < 0.05/8$ with Bonferroni correction for multiple comparison) symptoms.

Among all participants' (PCC, UPC, HNC) observations, sex-males could be identified with significant model performance with the cross-validated balanced accuracy (%) 81.786 ± 6.209 ($P$ = 0.001; $P < 0.05/8$ with Bonferroni correction for multiple comparisons), but PCC group cannot be identified with significance performance (55.425 ± 5.766; P = 0.125; P > 0.05). This study limits the model training to identify PCC vs non-PCC due to practical demands. The identification of other classes (e.g., UPC vs. non-UPC) or the implementation of multiclass classification (e.g., PCC vs. UPC vs. HNC) was not computed in this study due to a lack of practicality or clinical urgency.

## 3.2. Model explainability

The SVC coefficient calculates the magnitude of a feature that influences the decision boundaries between two classes, thus providing an opportunity to evaluate how the model derives its decisions by comparing the average SVC coefficient from the actual label against the null distribution of the SVC coefficient obtained from shuffling the target label one thousand times (Vieira et al., 2020). The alpha level 0.05 threshold of the permutation test allows us to prioritize our focus on the influential features rather than the massive amount of non-influential features; for target labels 'word finding difficulties', 'multitasking', and 'male' are provided in Figures 5, 6, and 7, respectively. Essential column features for target labels 'acute olfactory symptoms', 'chronic olfactory symptoms', 'fatigue', and 'attention/memory/PCC' were provided in Supplementary Figures S2 to S5, respectively. The feature coefficient for 'PCC' prediction from all participants is not provided, as the model's balanced accuracy does not exceed the alpha threshold of 0.05, making the contribution of the features unclear.

## 4. Discussion

The current analysis aims to identify post-COVID-19 symptoms and categorize participants into corresponding groups using a data-driven approach. Due to the characteristics of the dataset, post-COVID conditions are determined from pooled PCC and UPC participants, while sex-male is identified from all participants (Table 1, Figures 1 and 2). The PCC symptoms, which consist of 'word finding difficulties' and 'multitasking', can be identified with model performance exceeding the alpha level of 0.05/8 for multiple tests (Figure 4) from PCC and UPC participants. Furthermore, the sex-male can be identified from the pooled PCC, UPC, and HNC participants with model performance exceeding the alpha level of 0.05/8 for multiple tests (Figure 4). To the author's knowledge, this is the first study to identify the occurrence of post-COVID-19 symptoms based on brain structural MRI input features.

Long-COVID symptoms may appear as a "brain fog," which refers to cognitive disturbances that encompass symptoms including the occurrence of 'word finding difficulties', 'multitasking', 'attention', and 'memory' disturbances (Pan et al., 2024; Yu and Absar, 2022). A case report with 5-month follow-up of a COVID patient with 'word-finding difficulties' complaints, by utilizing brain MRI and $^{18}$fluorodeoxyglucose-positron emission tomography (F-FDG-PET), described abnormality within the parietal lobe, left temporal lobe, and left frontal regions (Yu and Absar, 2022). Despite the difference in imaging modality and number of participants, the SVC coefficient also highlights the left postcentral sulcus, left transverse temporal sulcus, left parahippocampal gyrus, and left middle frontal sulcus as influential regions (Figure 5). In a case report of a patient with worsening multitasking ability, a brain MRI abnormality was identified in the right and left parietal regions (Bhaiyat et al., 2022); the SVC coefficient also identifies the left post-central sulcus as an influential region. A cross-sectional study of 86 participants with subjective cognitive complaints following SARS-CoV-2 infection and 36 healthy controls found hyperconnectivity between the left and right parahippocampal areas (Díez-Cirarda et al., 2023); the SVC coefficient also identified the left parahippocampal gyrus as the influential feature (Figure 6). A systematic review of patients with post-COVID-19 memory disturbances (Shan et al., 2022), reports functional abnormalities in the frontal, parietal, and temporal regions. In addition, a cross-sectional study of 24 patients with a non-invasive Arterial Spin Labeling (ASL)

MRI technique analysis found significant hypoperfusion also in the frontal, parietal, and temporal areas (Ajčević et al., 2023); the SVC coefficient highlights several influential regions that also correspond to those areas (Supplementary Figure S2).

Among COVID-19 patients, several brain regions have been associated with fatigue. For example, a cross-sectional study of 46 COVID-19 survivors and 30 controls using voxel-based morphometry on T1-weighted MRI images identified the posterior cingulate, precuneus, and superior parietal lobule as the most affected regions (Hafiz et al., 2022). Another cross-sectional study, involving 30 patients with post-COVID fatigue syndrome and 20 healthy volunteers, that underwent a functional MRI (fMRI) scan, identified the supramarginal gyri, opercular parts of the precentral gyri, and the posterior lobe of the cerebellum (Tanashyan et al., 2024). Another cross-sectional study of 56 COVID-19 patients and 37 matched controls, using quantitative diffusion-weighted MRI (d-MRI), identified the corpus callosum, arcuate fasciculus, cingulate gyrus, and fornix as affected areas (Bispo et al., 2022). A multi-center, longitudinal study of 79 post-COVID-19 patients with persistent symptoms and 21 healthy controls underwent T1-weighted anatomical scans and functional imaging sequences, which identify the pre- and postcentral gyrus and the limbic olfactory network association with fatigue (Dadsena et al., 2025). Accordingly, the SVC coefficient result also identifies the right superior parietal lobule, left precuneus, right supramarginal gyrus, left postcentral sulcus, and right anterior part of the cingulate gyrus and sulcus as influential regions (Supplementary Figure 3).

In addition to cognitive symptoms, COVID-19 can also cause acute and chronic olfactory symptoms. A case-control study of 15 participants with affected olfactory symptoms and 5 participants without symptoms who underwent functional MRI scanning found that the upper frontal lobe and basal ganglia are associated with olfactory dysfunction in COVID-19 patients (Iravani et al., 2024). A literature study also highlights the association between the entorhinal cortex and olfactory dysfunction in various neurodegenerative diseases (Barresi et al., 2012). Accordingly, the SVC coefficient indicates an association between left and right caudate volumes and acute olfactory symptoms, as well as an association between the left entorhinal cortical surface and chronic olfactory symptoms (Supplementary Figures 4 and 5). Considering that the neurodegenerative disease also shares a common affected region with post-COVID-19 patients (e.g., the entorhinal cortex), the changes in the corresponding brain region are unlikely to be caused solely by post-COVID-19 infection. Therefore, it is more reasonable to utilize brain MRI for identifying the occurrence of a specific symptom (e.g., olfactory deficit) instead of the particular cause (e.g., post-COVID-19 infection). Which symptoms are the most recognizable by brain MRI features will need a further exploration.

In addition, the Freiburg-MRI tabular dataset is also capable of predicting the sex of males, with model performance exceeding other predictions for long-COVID symptoms and PCC group participants. Giedd et al. reviewed that the caudate nucleus, amygdala, hippocampus, and cerebellum are structures that commonly exhibit differences between genders (Giedd et al., 2012); the SVC coefficient also identifies the cerebellum and the parahippocampal gyrus as influential features (Figure 7). Although sex-male identification is not the primary purpose of the current analysis, the replicability of previous findings (Baecker et al., 2021) reflects the quality of the data recordings provided by the Freiburg-MRI dataset, which also contributes to our current understanding of the role of brain MRI in sex prediction.

In addition to influential feature identifications, the different target label shows various performances; for example, 'word finding difficulties' and 'multitasking' show a higher balanced accuracy than 'PCC' when the model tries to identify post-COVID symptoms among the PCC and UPC participants. Furthermore, predicting 'PCC' participants by using all participants shows balanced accuracy performance that does not exceed the chance level (Figure 4). Since the 'PCC' population accumulates various symptoms from individuals, it supports the argument that the brain MRI features may have better separability when it comes to specific symptoms as a target label rather than targeting the 'PCC' group in general. This finding may benefit further research aimed at identifying particular PCC symptoms (as opposed to the overall PCC group) using neuroimaging modalities.

The primary research group has reanalyzed the brain MRI features from the same PCC and UPC participants that was also utilized in this manuscript analysis. By utilizing more features derived from the structural brain MRI (116 Tissue probability values (TPV)-derived features, 912 features obtained by Diffusion tensor imaging (DTI), 684 by neurite orientation dispersion and density imaging (NODDI), and 684 by diffusion microstructure imaging (DMI), feature selection method, and linear SVM, the PCC and UPC could be decoded with ROC AUC between 0.59 – 0.95 (min - max) (Rau et al., 2025). While Rau et al.'s recent work focused on utilizing machine learning to predict the broad Post-COVID Condition (PCC) diagnosis, our primary analysis focuses on identifying the specific MRI features that predict individual long COVID symptoms, such as fatigue and word-finding difficulties. Nevertheless, decoding of PCC from limited UPC-PCC participants and all participants serves as a comparative extension of initial intentions. Although there is a difference in available features, algorithm, and cross-validation implementation, PCC and UPC decoding based on the anonymized public dataset show ROC AUC of 0.69 (0.07) (mean (standard deviation)) (See Supplementary Table S1), which is lower than the maximum ROC AUC finding reported by Rau et al., (Rau et al., 2025), but still within the range of their findings. This opens further possibilities to improve the PCC symptoms model performance by incorporating reasonable features and advanced machine learning techniques.

This current study has several limitations. As an observational study, overlapping symptoms among participants are unavoidable. These overlapping symptoms may occur partially or entirely (see Supplementary Figure S1 for an UpsetPlot of various label intersections among participants), thus resulting in shared brain regions identified by the support vector coefficient across independent models. For example, the left insula (Figures 4 and 5, Supplementary Figures 2 and 3) is recognized as the influential feature in separated models predicting word-finding difficulties, multitasking, attention, and fatigue symptoms. The non-mutually-exclusive labelling has limited further interpretation, whether the identified common brain areas are responsible for those multiple symptoms simultaneously, or whether the shared brain areas are actually involved in only one symptom but are entangled by the overlapping symptoms experienced by the patients.

The Freiburg-MRI dataset has contributed to a massive collection of hundreds of anonymized participants with corresponding brain features and post-COVID-19 symptoms that have been made public. However, there are challenges due to the "more features than observations" and the limited availability of participants for training the between-subject model; thus, the current study employs SVC algorithms and focuses on the binary classification task, despite the availability of ordinal labels. ML classification algorithms are exceptionally reliable for predicting binary targets, even with a small number of observations (Hernández et al., 2014; Khondoker et al., 2016). There are empty labels, especially among the HNC group, which limit the model's prediction of post-COVID symptoms to only pooled PCC and UPC participants (in contrast to the sex-male prediction that can be made across all participants). It also has some imbalanced labels, depending on the selected target. However, the nested cross-validation for model development, with balanced accuracy as the target metric and a permutation test as the significance test procedure, has been employed to minimize bias in deriving conclusions under such circumstances. Since external validation data is not available, this current work mainly served as a 'proof-of-concept' rather than a 'ready-to-implement' study. The explainable ML performance does not imply causality, but the identified region could open further analysis to determine causality. This current analysis does not employ a variety of ML or a combination with complex additional algorithms to derive a conclusion from the data itself; further analysis with more varied and complex algorithms could be implemented for future studies. Furthermore, although the long-COVID symptoms encompass multiple organs (Unger, 2025), the implementation of brain structural MRI to identify the long-COVID symptoms should be limited to relevant symptoms of the brain network.

Overall, this study demonstrated the benefit of incorporating brain structural MRI for identifying long-COVID symptoms, with potential applications for patient screening. The brain structural MRI provides an objective measure of brain microstructure, thus serving as a neuroimaging-based biomarker candidate. It is less dependent on the patient's report; therefore, it could give a lead to possible long-COVID symptoms occurrence when conventional diagnosis was stagnated due to unrecorded prior mild symptoms, a false-negative RT-PCR result, and difficulties in communicating the patient's symptoms. The explainable ML ability to identify associated regions also serves to expand our current knowledge about the potential involved areas in PCC patients. In the future, it is expected that a larger dataset from multiple centers representing diverse patient demographics, with a separate external validation group, and a uniform labeling procedure, could provide a model that is ready for deployment.

In conclusion, brain structural MRI could identify long-COVID symptoms and the sex of the participant. The ML task of predicting specific long-COVID symptoms is more favorable than predicting the PCC group, either solely from the COVID-19 survivors or from all participants. The brain structural MRI could potentially serve as an alternative modality for identifying the occurrence of long-COVID symptoms, pointing out the associated region affected after COVID-19 infection, and reducing the physician's dependence on patient reports.

**Acknowledgements**

The author (AR) acknowledges University Hospital Freiburg as the monocentric research location, and the authors of source manuscript that made the research and anonymized dataset accessible for public (Jonas A. Hosp, Marco Reisert, Andrea Dressing, Veronika Götz, Elias Kellner, Hansjörg Mast, Susan Arndt, Cornelius F. Waller, Dirk Wagner, Siegbert Rieg, Horst Urbach, Cornelius Weiller, Nils Schröter & Alexander Rau); Projekt DEAL that supported Open Access funding, Berta-Ottenstein-Programme for Clinician and Advanced Clinician Scientists for supporting the authors of source research. The author thanks all participants that has contributed to the primary research and dataset collection.

**Data availability**

The anonymized dataset can be accessed from the Dryad data center (https://doi.org/10.5061/dryad.kkwh70s9g).

**Ethics statement**

Ethical review for the primary study was approved by the Ethics Committee of the University of Freiburg (EK 211/20). The study was registered in the 'Deutsches Register Klinischer Studien (DRKS)' (DRKS00021439), and all participants provided written informed consent for their participation in the study. The datasets used for analysis in this study are anonymized and publicly accessible on a public repository (https://doi.org/10.5061/dryad.kkwh70s9g). Due to the implementation of secondary analysis from an anonymized dataset, which is made accessible and distributed under a public domain license (https://creativecommons.org/publicdomain/zero/1.0/), this current study does not require further IRB approval.

Table 1. Participants demographics

| Variables | Presence | PCC (*n* = 89) | UPC (*n* = 38) | HNC (n = 46) |
|---|---|---|---|---|
| Male | No | 55 (61.80%) | 25 (65.79%) | 23 (50.00%) |
| Male | Yes | 34 (38.20%) | 13 (34.21%) | 23 (50.00%) |
| acute olfactory symptoms | No | 23 (25.84%) | 26 (68.42%) | - |
| acute olfactory symptoms | Yes | 66 (74.16%) | 12 (31.58%) | - |
| attention symptoms | No | - | 38 (100.00%) | - |
| attention symptoms | Yes | 89 (100.00%) | - | - |
| chronic olfactory symptoms | No | 45 (50.56%) | 38 (100.00%) | - |
| chronic olfactory symptoms | Yes | 44 (49.44%) | - | - |
| fatigue symptoms | No | 4 (4.49%) | 38 (100.00%) | - |
| fatigue symptoms | Yes | 85 (95.51%) | - | - |
| memory symptoms | No | - | 38 (100.00%) | - |
| memory symptoms | Yes | 89 (100.00%) | - | - |
| multitasking symptoms | No | 3 (3.37%) | 38 (100.00%) | - |
| multitasking symptoms | Yes | 86 (96.63%) | - | - |
| word finding difficulties | No | 10 (11.24%) | 38 (100.00%) | - |
| word finding difficulties | Yes | 79 (88.76%) | - | - |

Table 1. Participants demographics. Column variables represent phenotype and post-COVID-19 symptoms. The values outside the brackets under column PCC, UPC, and HNC represent the count of phenotype and symptom occurrence as 'Yes' or 'No', and the values inside the brackets represent percentage (%). PCC: the Post-COVID-19 Condition , UPC: the Unimpaired Post-COVID group, HNC: the Healthy Controls group.

Figure 1. Participants inclusion flowchart

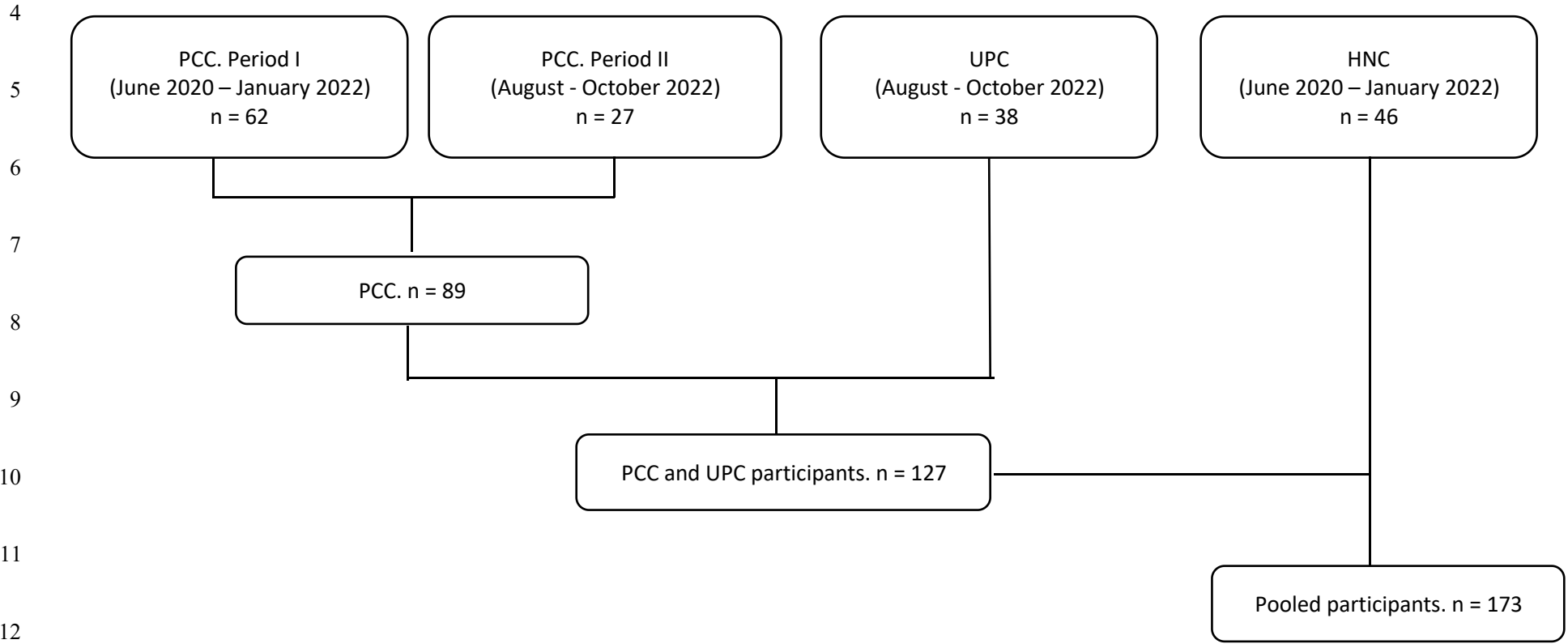

Figure 1. Participants' inclusion flowchart. Participants are recruited into three separate groups: the Post-COVID-19 Condition (PCC) group, the Unimpaired Post-COVID group (UPC), and the Healthy Controls group (HNC).

Figure 2. Overview of model development.

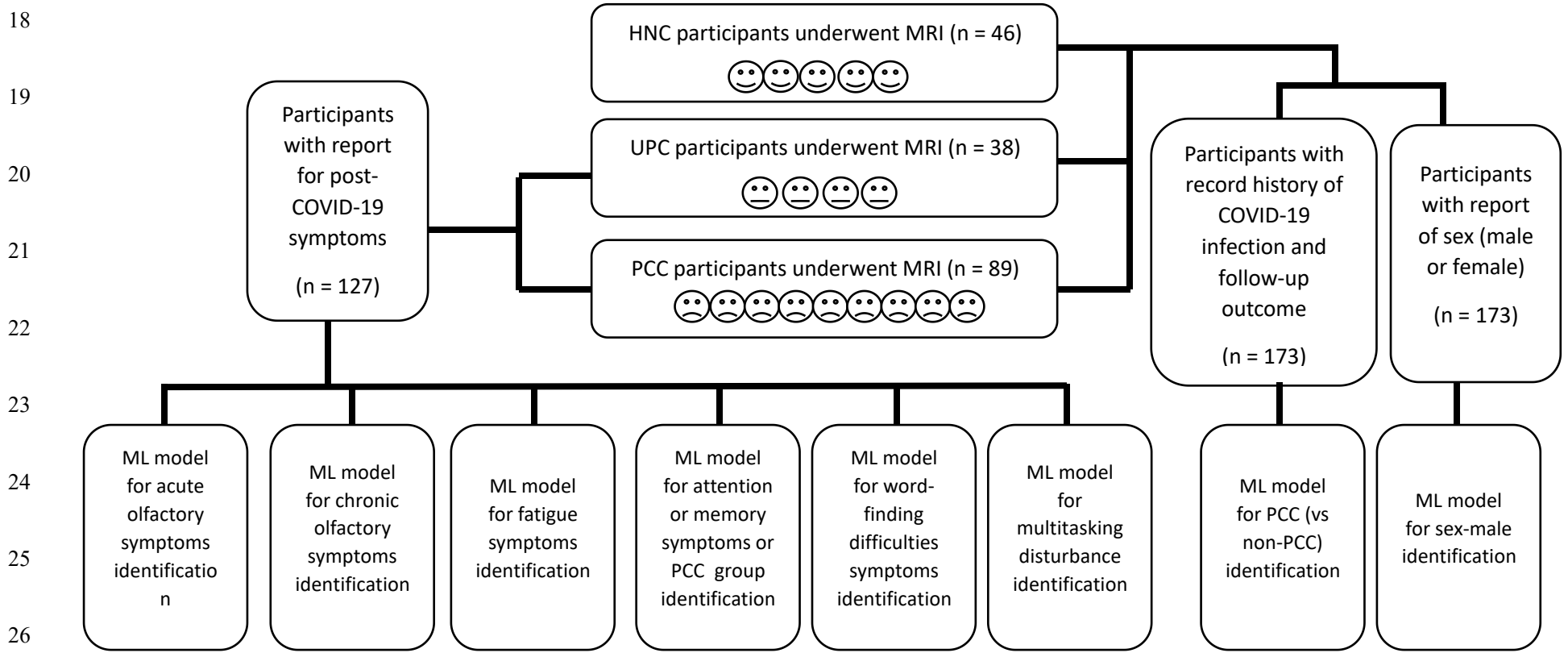

Figure 2. Overview of model development. The brain MRI features from HNC, UPC, and PCC participants with information about sex and their corresponding group were prepared for independent model training for the identification of PCC vs non-PCC and sex-male. The brain MRI features from UPC and PCC participants, along with information about post-COVID-19 symptoms, were prepared for independent model training to identify acute and chronic olfactory symptoms, fatigue, attention/memory-symptoms/PCC-group, word-finding difficulties, and multitasking disturbances. HNC: the Healthy Controls group, PCC: the Post-COVID-19 Condition, UPC: the Unimpaired Post-COVID group. ML: machine learning. MRI: magnetic resonance imaging.

Figure 3. Phi coefficient of PCC symptoms occurrence among PCC and UPC participants

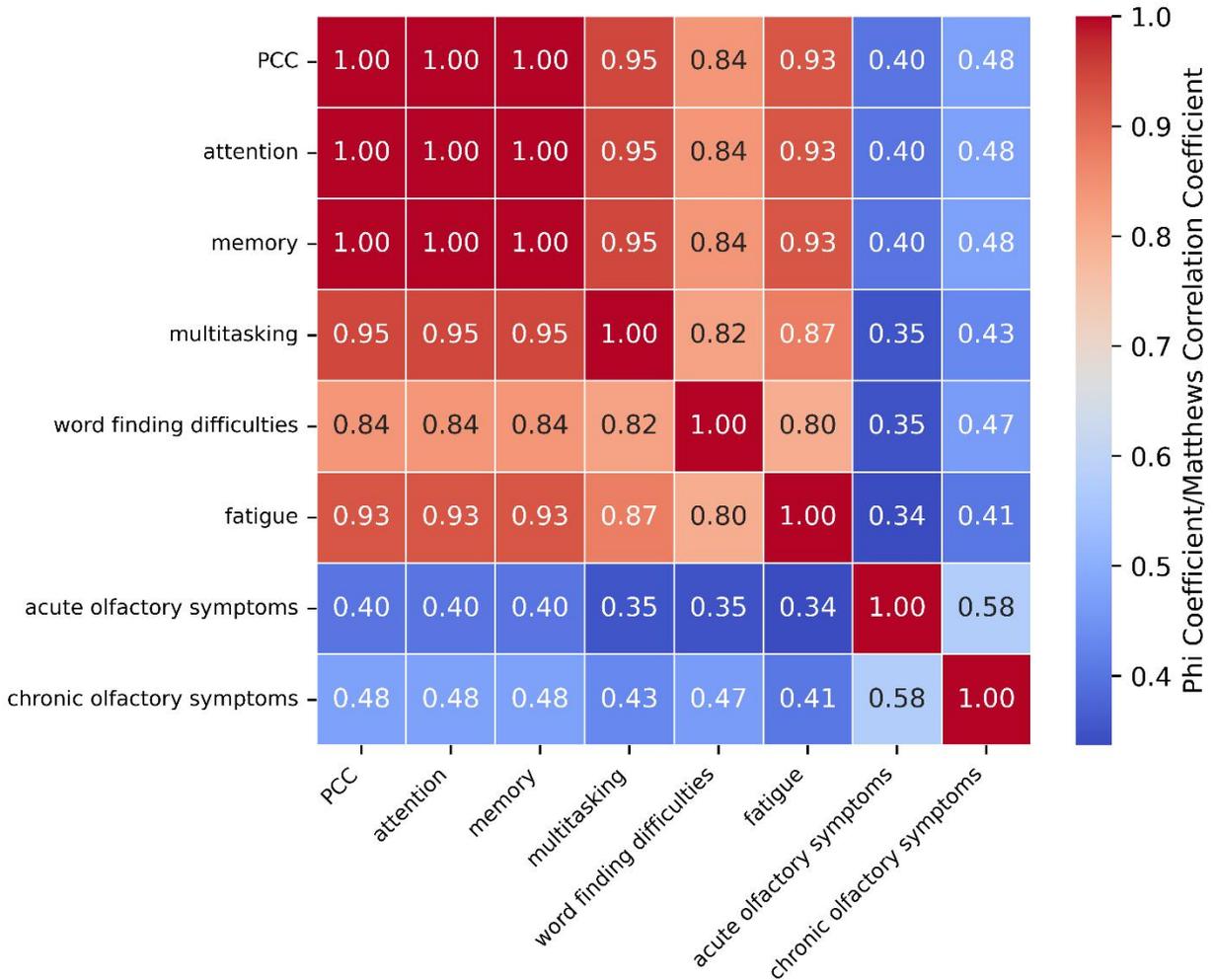

Figure 3. Matthew's correlation coefficient (MCC) between binary variables among pooled PCC and UPC participants.  It describes the binary correlation across the target label among the PCC and UPC groups, and all PCC participants reported attention and memory deficits. PCC: the Post-COVID-19 Condition, UPC: the Unimpaired Post-COVID group, HNC: the Healthy Controls group.

Figure 4. Support vector classifier model performance across various target labels compared to one thousand permutations of the label.

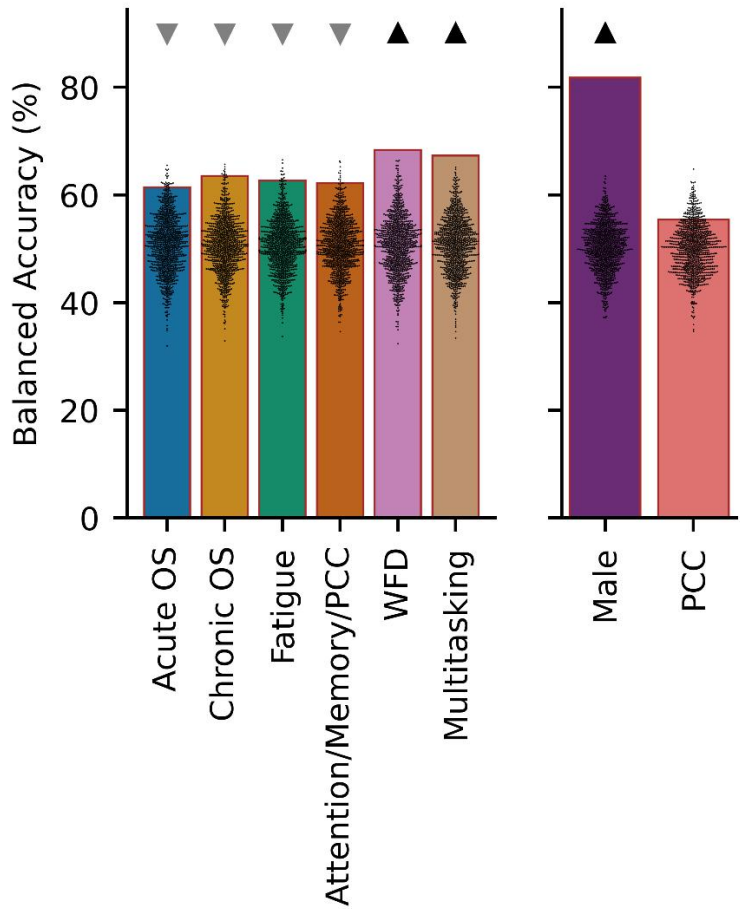

Figure 4. SVC model performance across different labels and different pooled participants. The left-side partition figure represents PCC symptom prediction among pooled PCC and UPC participants. The right side partition figure represents the PCC group and the male prediction among the pooled PCC, UPC, and HNC participants. The bar plot represents the average balanced accuracy (%) across outer loop nested cross-validation results. The black dot inside the swarm plot represents the average balanced accuracy (%) of outer-loop nested cross-validation from each of the unique shuffling label results, obtained from 1000 iterations. The inverted grey triangle represents $P < 0.05$, and the black triangle represents $P < 0.05/8$. OS: olfactory symptoms, PCC: post-COVID-19 conditions, WFD: word-finding difficulties.

Figure 5. Word finding difficulties prediction's SVC coefficients.

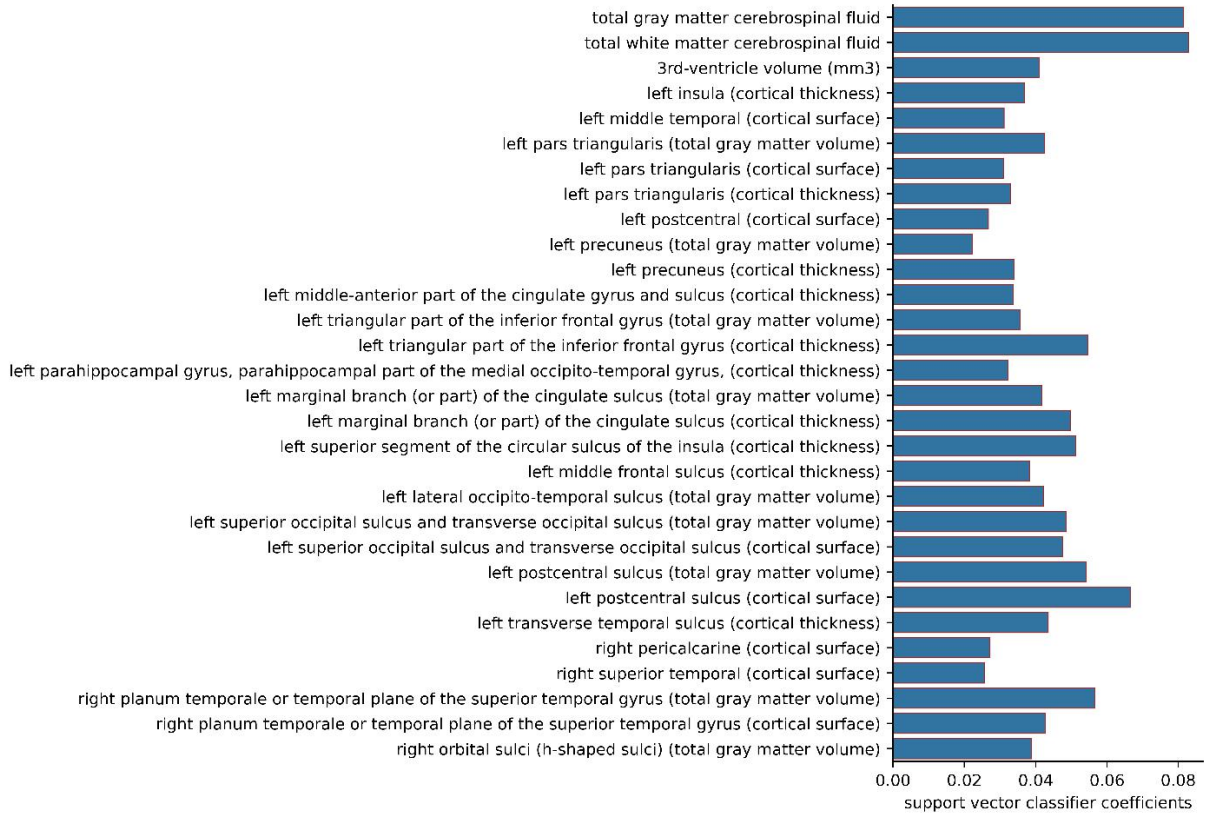

Figure 5. Horizontal barplot of the SVC coefficient from the SVC model that identifies the occurrence of word-finding-difficulties symptoms. The x-axis represents SVC coefficient values. The y-axis represents feature columns that exceed the alpha threshold of 0.05.

Figure 6. Multitasking prediction's SVC coefficients.

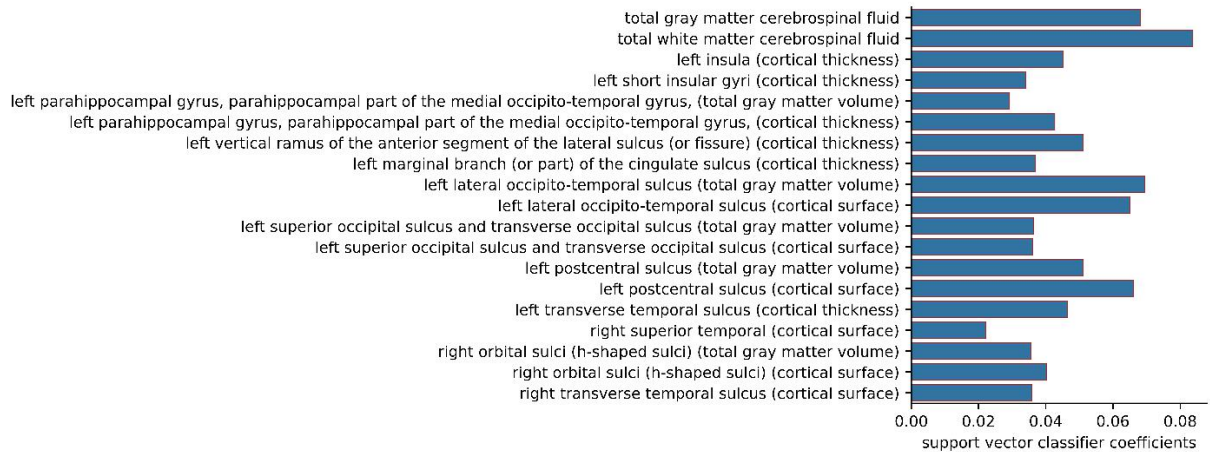

Figure 6. Horizontal barplot of the SVC coefficient from the SVC model that identifies the occurrence of multitasking difficulties. The x-axis represents SVC coefficient values. The y-axis represents feature columns that exceed the alpha threshold of 0.05.

Figure 7. Male prediction's SVC coefficients

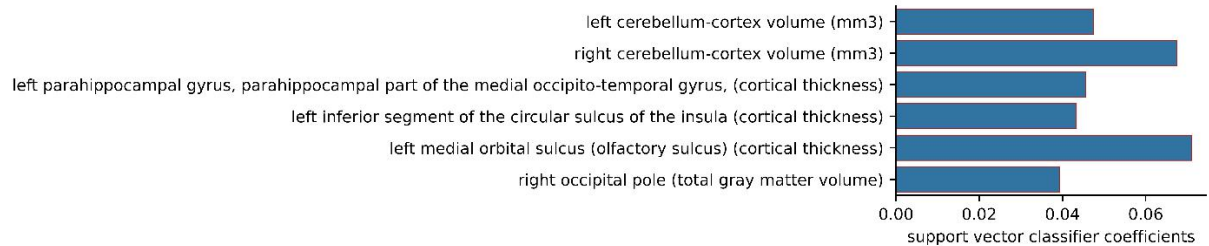

Figure 7. Horizontal barplot of the SVC coefficient from the SVC model that identifies the sex-male. The x-axis represents SVC coefficient values. The y-axis represents feature columns that exceed the alpha threshold of 0.05.

Supplementary Materials

Supplementary Figure S1

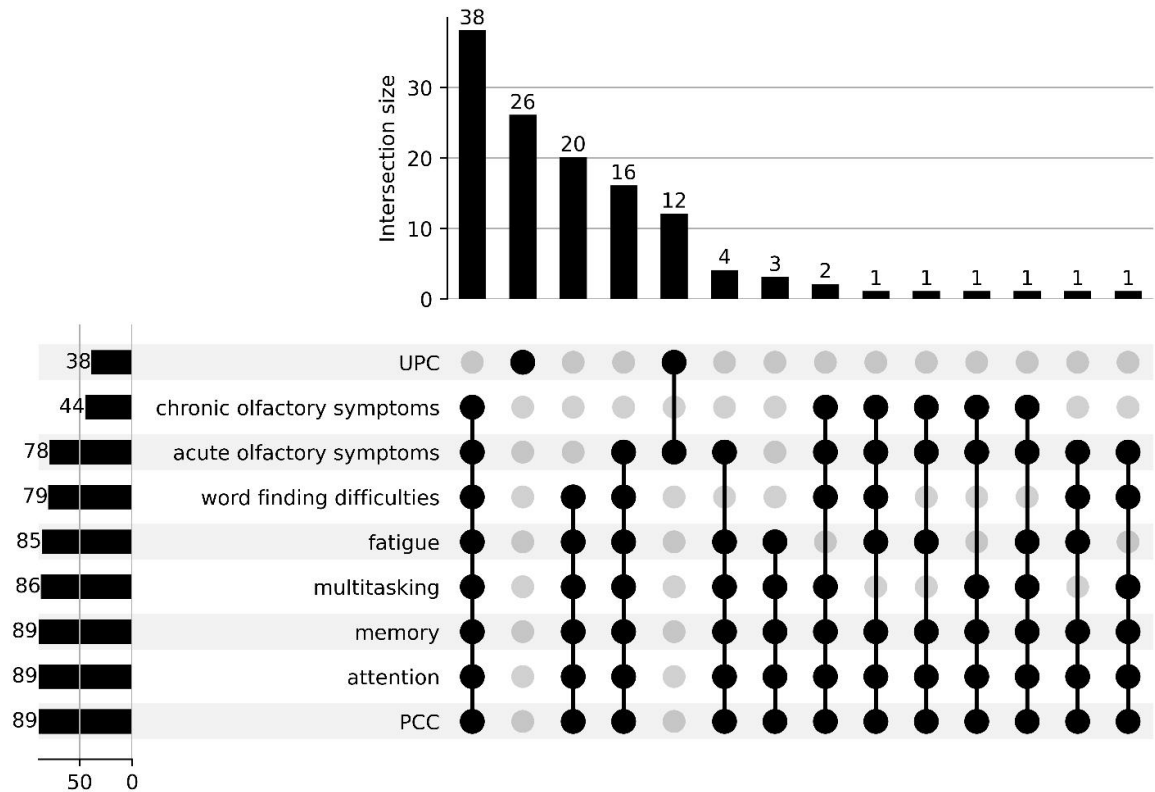

Supplementary Figure S1. Upsetplot representing intersections between variables among participants. Connected circles representing intersections among corresponding variables. Vertical barplot repsenting count from corresponding intersections between variables. Horizontal barplot representing count from corresponding variables.

Supplementary Figure S2

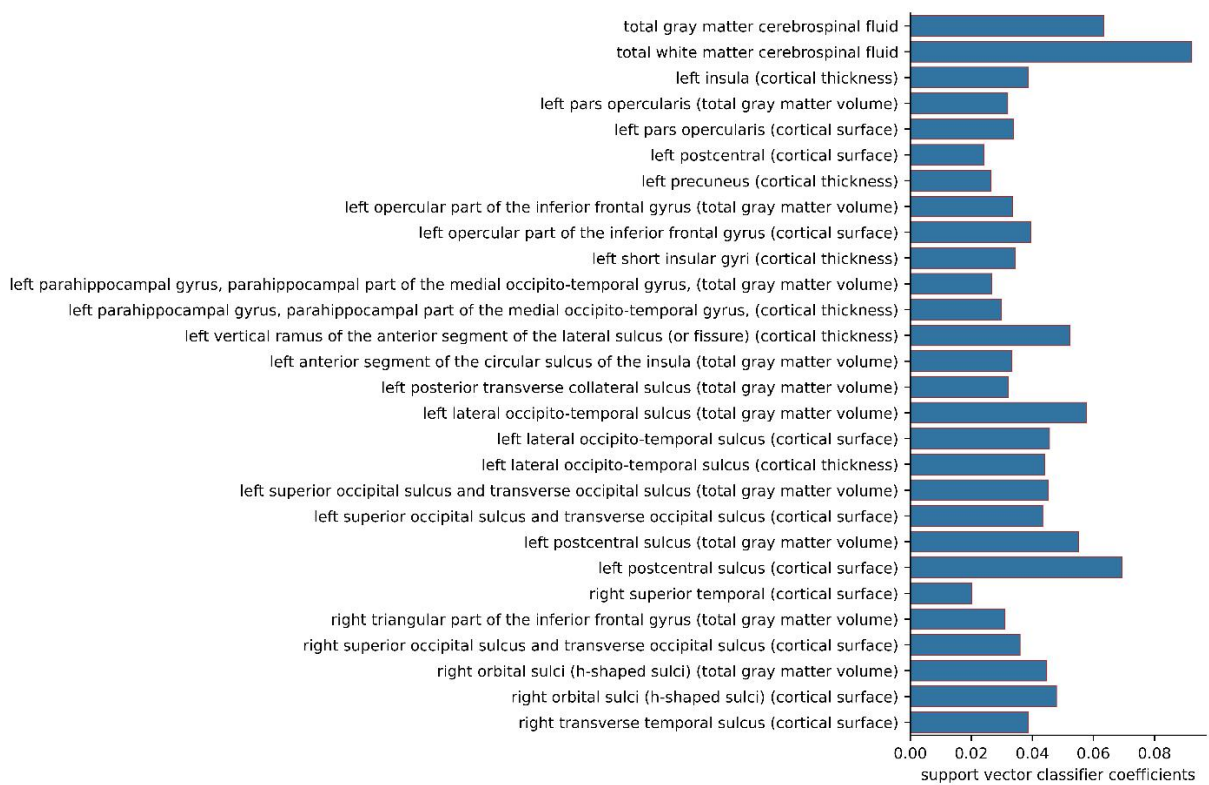

Supplementary Figure 2. Horizontal barplot of the SVC coefficient from the SVC model that identifies the occurrence of attention/memory disturbances/PCC group. The x-axis represents SVC coefficient values. The y-axis represents feature columns that exceed the alpha threshold of 0.05. PCC: Post COVID-19 conditions.

Supplementary Figure S3

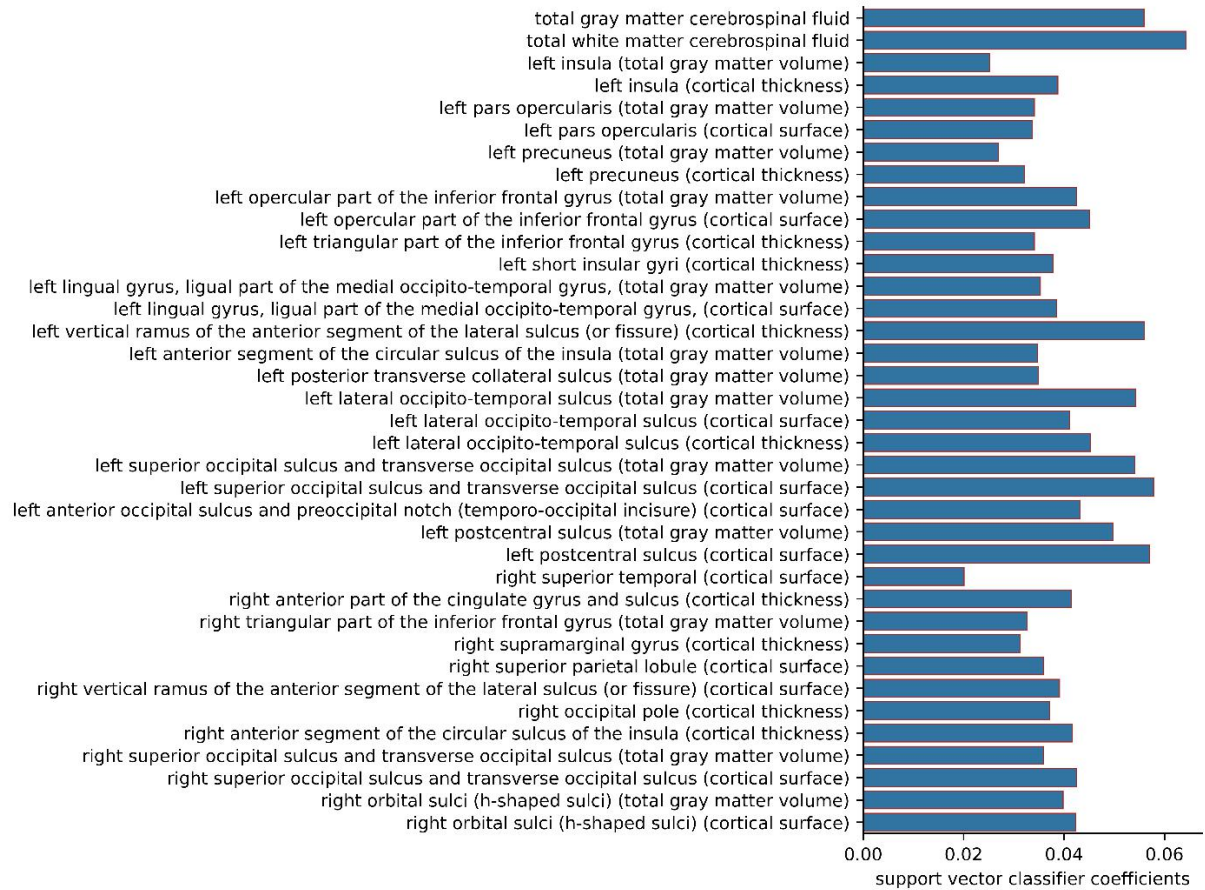

Supplementary Figure 3. Horizontal barplot of the SVC coefficient from the SVC model that identifies the occurrence of fatigue. The x-axis represents SVC coefficient values. The y-axis represents feature columns that exceed the alpha threshold of 0.05.

Supplementary Figure S4

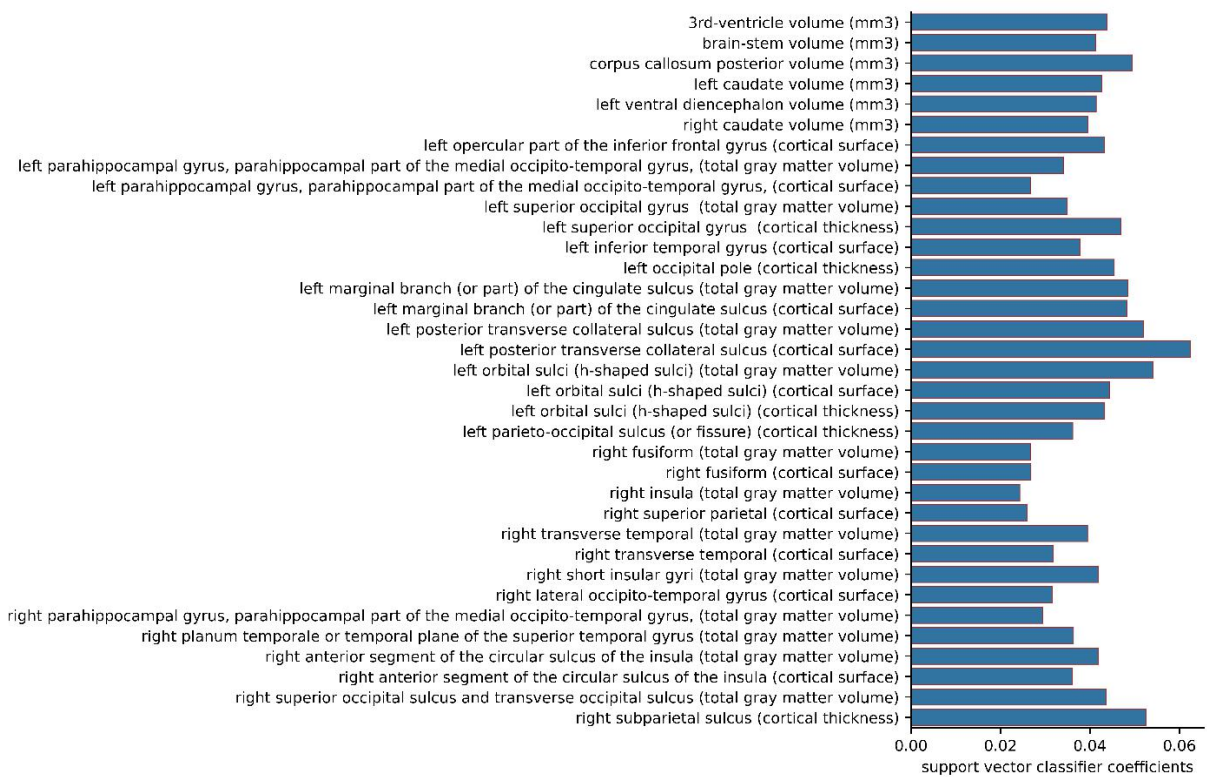

Supplementary Figure 4. Horizontal barplot of the SVC coefficient from the SVC model that identifies the occurrence of acute olfactory symptoms. The x-axis represents SVC coefficient values. The y-axis represents feature columns that exceed the alpha threshold of 0.05.

Supplementary Figure S5

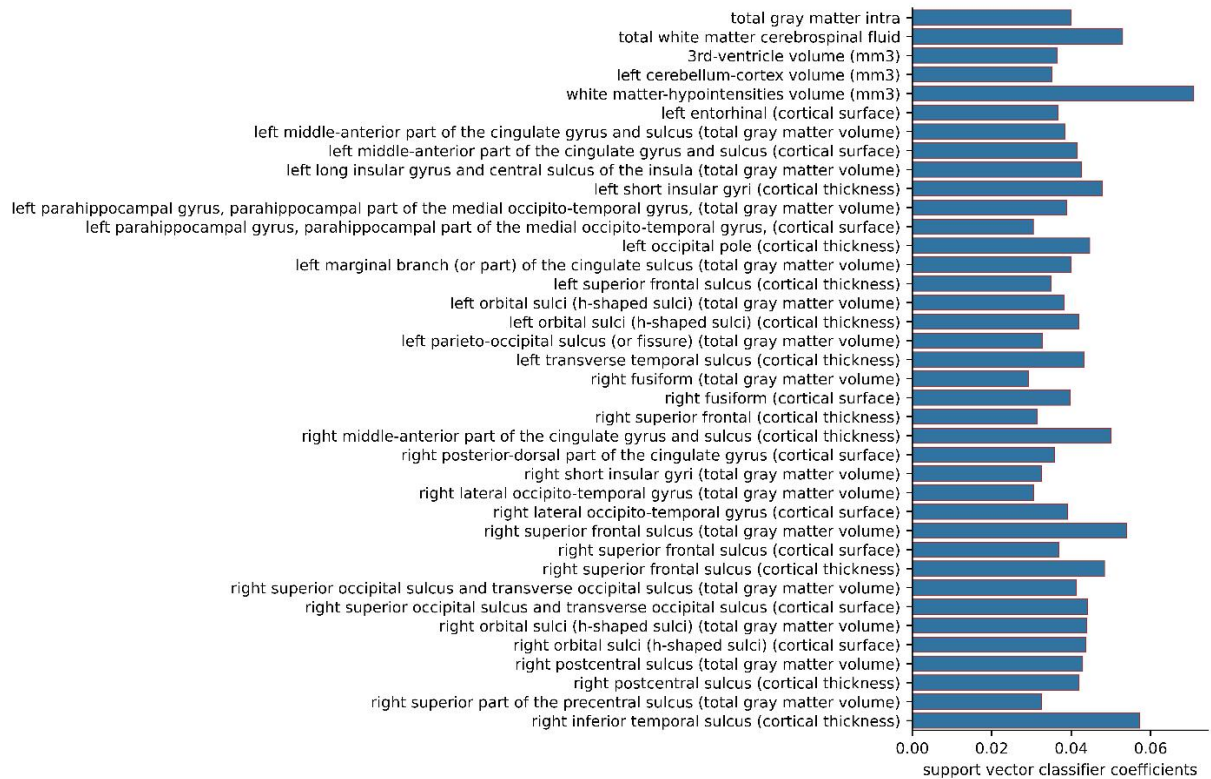

Supplementary Figure 5. Horizontal barplot of the SVC coefficient from the SVC model that identifies the occurrence of chronic olfactory symptoms. The x-axis represents SVC coefficient values. The y-axis represents feature columns that exceed the alpha threshold of 0.05.

Supplementary Table S1

| Included population | Target label | Specificity | Negative predictive value | Balanced accuracy | Accuracy | Mathew coefficient | F1 score | Recall score | Precision score | Cohen kappa score | ROC AUC |
|---|---|---|---|---|---|---|---|---|---|---|---|
| PCC, UPC, HNC | Sex-male | 0.84 (0.10) | 0.86 (0.06) | 0.82 (0.06) | 0.82 (0.06) | 0.64 (0.13) | 0.78 (0.08) | 0.80 (0.11) | 0.78 (0.13) | 0.63 (0.13) | 0.90 (0.04) |
| PCC, UPC, HNC | PCC | 0.52 (0.15) | 0.53 (0.05) | 0.55 (0.06) | 0.55 (0.06) | 0.11 (0.12) | 0.58 (0.02) | 0.58 (0.04) | 0.57 (0.07) | 0.11 (0.11) | 0.56 (0.10) |
| PCC, UPC | word finding difficulties | 0.61 (0.16) | 0.63 (0.15) | 0.68 (0.09) | 0.70 (0.09) | 0.38 (0.18) | 0.76 (0.08) | 0.76 (0.13) | 0.76 (0.08) | 0.37 (0.18) | 0.72 (0.04) |
| PCC, UPC | memory | 0.41 (0.19) | 0.54 (0.28) | 0.62 (0.10) | 0.71 (0.08) | 0.27 (0.24) | 0.80 (0.06) | 0.83 (0.12) | 0.77 (0.04) | 0.26 (0.22) | 0.76 (0.04) |
| PCC, UPC | fatigue | 0.45 (0.16) | 0.53 (0.11) | 0.63 (0.08) | 0.69 (0.07) | 0.27 (0.16) | 0.77 (0.06) | 0.80 (0.10) | 0.75 (0.06) | 0.26 (0.16) | 0.64 (0.04) |
| PCC, UPC | multitasking | 0.56 (0.28) | 0.59 (0.21) | 0.67 (0.12) | 0.71 (0.11) | 0.37 (0.26) | 0.78 (0.10) | 0.78 (0.17) | 0.81 (0.13) | 0.34 (0.24) | 0.74 (0.06) |
| PCC, UPC | acute olfactory symptoms | 0.51 (0.19) | 0.56 (0.14) | 0.61 (0.06) | 0.64 (0.05) | 0.24 (0.13) | 0.71 (0.06) | 0.72 (0.15) | 0.71 (0.05) | 0.23 (0.12) | 0.67 (0.04) |
| PCC, UPC | chronic olfactory symptoms | 0.79 (0.13) | 0.75 (0.08) | 0.64 (0.11) | 0.69 (0.10) | 0.28 (0.23) | 0.49 (0.21) | 0.48 (0.24) | 0.54 (0.16) | 0.27 (0.23) | 0.69 (0.07) |

Supplementary Table S1. Report of other metrics derived from averaging the outer-loop nested CV performances. Values represented in decimal as scikit-learn output. Values outside the brackets represent the model performance's mean of the corresponding label and metrics; values inside the brackets represent the standard deviation. ROC AUC: receiver operating characteristics – area under the curve. PCC: the Post-COVID-19 Condition , UPC: the Unimpaired Post-COVID group, HNC: the Healthy Controls group.